\documentclass[twocolumn,twocolappendix]{aastex631}

\usepackage{float}
\usepackage{multirow}
\usepackage{threeparttable}
\usepackage{tabularx}
\usepackage{amsmath}
\usepackage{hyperref}

\shorttitle{Automatic Astrometry and Photometry with Imfit }
\shortauthors{Amador-Portes et al.}
\begin{document}

\title{Automated Modeling with AAP-Imfit: Astrometry and Photometry via CASA}

\correspondingauthor{Alfredo Amador-Portes}
\email{alfre\_portess97@hotmail.com, aamador@inaoe.com}
\repository{GitHub repository on: \url{https://github.com/Alfred97AstroAGN/AAP-Imfit-aCASA-tool.git}}

\author[0009-0009-6341-0270]{Alfredo Amador-Portes}
\affiliation{Instituto Nacional de Astrofísica, Óptica y Electrónica, Luis Enrique Erro \#1, Tonantzintla Puebla, México, C.P. 72840}

\author[0009-0000-0630-2171]{Eva Palafox}
\affiliation{Instituto Nacional de Astrofísica, Óptica y Electrónica, Luis Enrique Erro \#1, Tonantzintla Puebla, México, C.P. 72840}
\affiliation{Max-Planck-Institut für Radioastronomie, Auf dem Hügel 69, D-53121 Bonn, Germany}

\author[0000-0002-5442-818X]{Víctor M. Patiño-Álvarez}
\affiliation{Instituto Nacional de Astrofísica, Óptica y Electrónica, Luis Enrique Erro \#1, Tonantzintla Puebla, México, C.P. 72840}
\affiliation{Max-Planck-Institut für Radioastronomie, Auf dem Hügel 69, D-53121 Bonn, Germany}

\author[0000-0002-2558-0967]{Vahram Chavushyan}
\affiliation{Instituto Nacional de Astrofísica, Óptica y Electrónica, Luis Enrique Erro \#1, Tonantzintla Puebla, México, C.P. 72840}

\author[0000-0003-1622-1484]{Andrei P. Lobanov}
\affiliation{Max-Planck-Institut für Radioastronomie, Auf dem Hügel 69, D-53121 Bonn, Germany}

\author[0000-0001-6010-6200]{Sergio A. Dzib}
\affiliation{Max-Planck-Institut für Radioastronomie, Auf dem Hügel 69, D-53121 Bonn, Germany}

\begin{abstract}
Very Long Baseline Interferometry (VLBI) provides the highest-resolution radio intensity maps, crucial for detailed studies of compact sources like active galactic nuclei (AGN) and their relativistic jets. Analyzing jet components in these maps traditionally involves manual Gaussian fitting, a time-consuming bottleneck for large datasets. To address this, we present an automated batch-processing tool, based on the Gaussian fitting capabilities of CASA, designed to streamline VLBI jet component characterization (AAP-Imfit). Our algorithm sets a detection limit, performs automatic 2D Gaussian fitting, and removes model artifacts, efficiently extracting component flux densities and positions. This method enables systematic and reproducible analysis, significantly reducing the time required for fitting extensive VLBI datasets. We validated AAP-Imfit by using VLBI observations of the blazars 3C 279 and 3C 454.3, comparing our results with published fits. The close agreement in residual root mean square (RMS) values and model/residual-to-map RMS ratios confirms the accuracy of our automated approach in reproducing original flux distributions. While visual inspection remains important for complex or faint features, this routine significantly accelerates VLBI component fitting, paving the way for large-scale statistical studies of jet dynamics.
\end{abstract}


\section{Introduction} \label{sec:intro}

Very Long Baseline Interferometry (VLBI) provides the highest angular resolution available in astronomy observations, enabling the study of compact radio sources such as active galactic nuclei (AGN) and their relativistic jets at parsec scales. The analysis of VLBI intensity maps is crucial for understanding the structural and flux evolution of these sources.  One of the major challenges in VLBI image characterization arises from the lack of well-defined edges between the source and the background in VLBI intensity maps. Unlike optical images, where it is often possible to clearly differentiate and subtract the background emission from the source, radio interferometric intensity maps frequently exhibit diffuse flux distributions that gradually blend into the background noise. This makes it difficult to define precise detection limits, leading to a potential misidentification of components or incorrect flux estimates. For example, \citet{Grobler2014} investigated calibration artifacts, called ghost sources, in radio interferometric data, highlighting how incomplete sky models can lead to such artifacts, complicating source characterization. Additionally, low-surface-brightness features can be challenging to differentiate from background fluctuations, particularly in jet regions where emission fades over distance. \citet{Condon2012} noted that the brightness distribution converges rapidly at micro Jansky levels, complicating the differentiation between discrete sources and background emissions. Similarly, \citet{Offringa2012} discussed the difficulty in separating off-axis sources from radio frequency interference due to the overlap of fringe frequencies in the uv-plane. As a result, an automated approach to estimating the flux distribution must incorporate robust noise estimation and adaptive detection limit to ensure that only statistically significant structures are fitted while avoiding overfitting of noise peaks or imaging artifacts.

Several algorithms have been developed to streamline the analysis of VLBI images. For instance, the wavelet-based image segmentation and evaluation method (WISE; \citealp{MertensAndLobanov2015}) allows for the detection and tracking of jet features across multiple epochs and scales but does not provide precise integrated flux measurements for individual components. SAND \citep{Zhang2018} offers an automated VLBI imaging and analysis pipeline, focusing on image reconstruction rather than component fitting. The difference of Gaussian wavelets and hard image thresholding (DoG-HiT; \citealp{Muller2022}) enhances VLBI de-convolution through multiscale modeling but does not optimize flux characterization for distinct jet features. More recently, deep learning techniques \citep{Lai2025} have been explored for VLBI image reconstruction, though their application to component-based flux extraction remains limited.

Traditionally, VLBI flux components analysis involves fitting 2-D Gaussians to jet features to characterize their astrometry (position) and photometry (flux density). This process allows us to track jet motion, variability, and identify key emission regions. However, the standard approach is often performed manually, on a map-to-map basis, which is time-consuming, and difficult to apply to large datasets. A widely adopted tool for VLBI component fitting is MODELFIT, part of the \texttt{DIFMAP} package \citep{Shepherd1997}, which fits circular Gaussian components directly in the uv-plane. This method benefits from working in the Fourier domain, avoiding image-based artifacts introduced by imaging and deconvolution procedures. This simplification reduces the number of free parameters and often yields stable fits for bright, well-isolated components. However, its use of circular Gaussians can be limiting: real jet features often exhibit elongated or asymmetric structures, and forcing circular symmetry can yield unphysical flux estimates especially in blended or regions with complex morphology. Such situations are common in parsec-scale jets, where emission features often exhibit elliptical or irregular morphologies due to projection effects and interactions within the jet. Moreover, VLBI restoring beams are typically elliptical, not circular, and ignoring this anisotropy can lead to biased flux estimates. A related approach is the cross-entropy global optimization (CE) technique introduced by \citet{Caproni2011}. This method models VLBI images using elliptical Gaussian components, optimizing their parameters through a stochastic global search algorithm. While the cross-entropy method provides accurate component fitting, particularly for complex jet structures, it is computationally expensive, being orders of magnitude slower than standard fitting techniques like the CASA task \texttt{imfit} \citep{Condon1997}, which employs a deterministic approach for Gaussian fitting.

We present a first approach to a fully automated open-source routine designed to fit 2-D Gaussian components to VLBI intensity maps, retrieving both astrometry and photometry for each fitted component, AAP-Imfit. This method systematically identifies jet features, estimates fluxes, and removes artifacts. This routine allows us to measure flux variability in jet components, along with their position. The methodology is particularly useful for long-term monitoring programs such as MOJAVE (Monitoring of Jets in Active Galactic nuclei with VLBA Experiments, \citealp{Lister2009}; \citeyear{Lister2013, Lister2018}) and BEAM-ME (Blazars Entering the Astrophysical Multi-Messenger Era, \citealp{Jorstad2017}), where various tens of VLBI maps must be analyzed consistently across multiple epochs. We describe the methodology used for the algorithm, including detection limit estimation, iterative component fitting, and artifact removal. We test the algorithm using VLBI maps of blazars 3C 279 and 3C 454.3, comparing our results with previous studies. A brief discussion of potential astrophysical applications is addressed, particularly for studying flux variability, Doppler boosting, and AGN jet dynamics. Finally, we summarize our conclusions and suggest potential improvements to the routine.

\begin{figure*}[t]
\centering
\includegraphics[width=2\columnwidth]{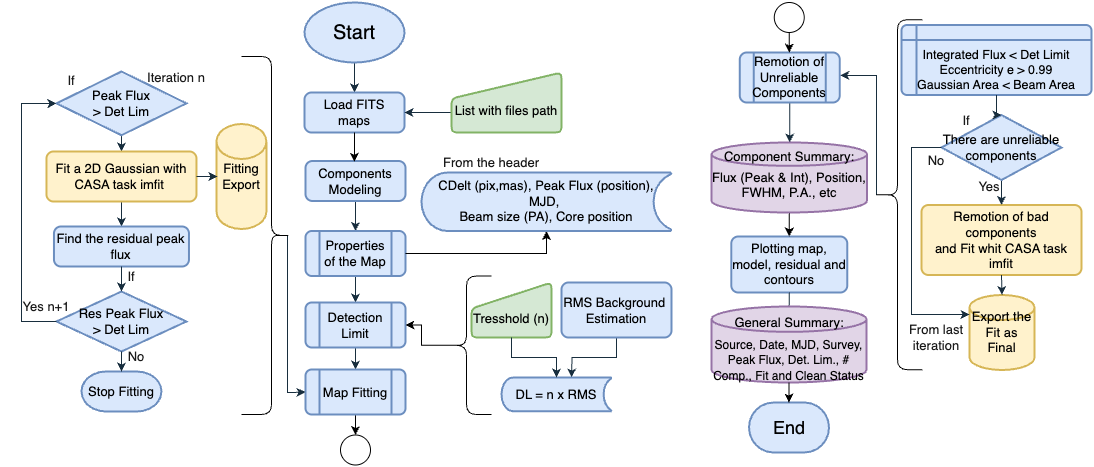}
\caption{Schematic of the AAP-Imfit routine. This flowchart is applied to each map to be modeled retrieving the astrometry and photometry of each Gaussian fitted. The algorithm consists of four main steps: First, all the parameters required by the code are extracted from the header of the \texttt{FITS}/\texttt{IMAP} file. Next, a detection limit is calculated, which separates real source emission from the background level. The final step involves removing Gaussian components that do not accurately represent the flux distribution. Finally, the map, model, and residuals are plotted for visualization, accompanied by a summary of the component properties and a general overview of all the fitted maps. Inputs and outputs are denoted in green and purple, respectively. Sections that use CASA tasks are denoted in yellow.}
\label{fig:flowchart}
\end{figure*}

\section{VLBI Astrometry and Photometry Fitting Algorithm }\label{sec:algorithm}

The Common Astronomy Software Applications package (CASA; \citealp{McMullin2007}) is a versatile software suite, widely used for processing and analyzing data from interferometers, such as VLBI intensity maps, and single-dish radio telescopes. The CASA tool \texttt{imfit} plays a key role in components characterization within astronomical images. Since \texttt{imfit} is designed to fit Gaussian components to emission regions in radio images, it provides detailed information about their position, size, orientation, and flux intensity. The fitting process in \texttt{imfit} employs a least-squares optimization method to determine the parameters of the Gaussian components, including the peak flux density, central coordinates, major and minor axes, and position angle. These parameters are returned alongside their respective uncertainties, which are derived from the covariance matrix of the fit\footnote{\texttt{imfit} documentation: \url{https://casadocs.readthedocs.io/en/latest/api/tt/casatasks.analysis.imfit.html}}. Further details can be found on \citet{McMullin2007}. Therefore, \texttt{imfit} can be used to characterize astrometry and photometry of individual components within a single intensity map.

As mentioned in \autoref{sec:intro}, algorithms of component characterization, e.g WISE or DoG-HiT, are capable of tracing components at different scales across multiple epochs, but without the ability to determine the integrated flux in each component. On the other hand, the CE optimization technique minimizes the difference between the observed and modeled images. For images with few components, the CE technique achieves accuracy comparable to traditional methods like \texttt{imfit}. But for larger number of Gaussian components the technique is computationally intensive, being substantially slower than traditional methods. Given this scenario, we developed an algorithm (AAP-Imfit) that relies on the CASA task \texttt{imfit} to identify structures in VLBI intensity maps to extract accurate astrometric (positions) and photometric (flux densities) information for these components. This is achieved through a systematic process involving noise estimation, thresholding, and iterative fitting. 

The directory where the \texttt{FITS/IMAP} maps are located must be provided by the user. The routine performs the following steps per map to model the flux distribution in the intensity maps:
\begin{enumerate}
\item Detection limit estimation.
\item Map fitting.
\item Remotion of unreliable components.
\item Data export and plotting of the results.
\end{enumerate}
In addition, a summary of the fitting status and statistics for all the epochs is exported. In \autoref{fig:flowchart} the basic outline of the algorithm is displayed. In the following subsections, each of the previous steps is described in detail.

\begin{figure*}[t]
\centering
\includegraphics[width=2.1\columnwidth]{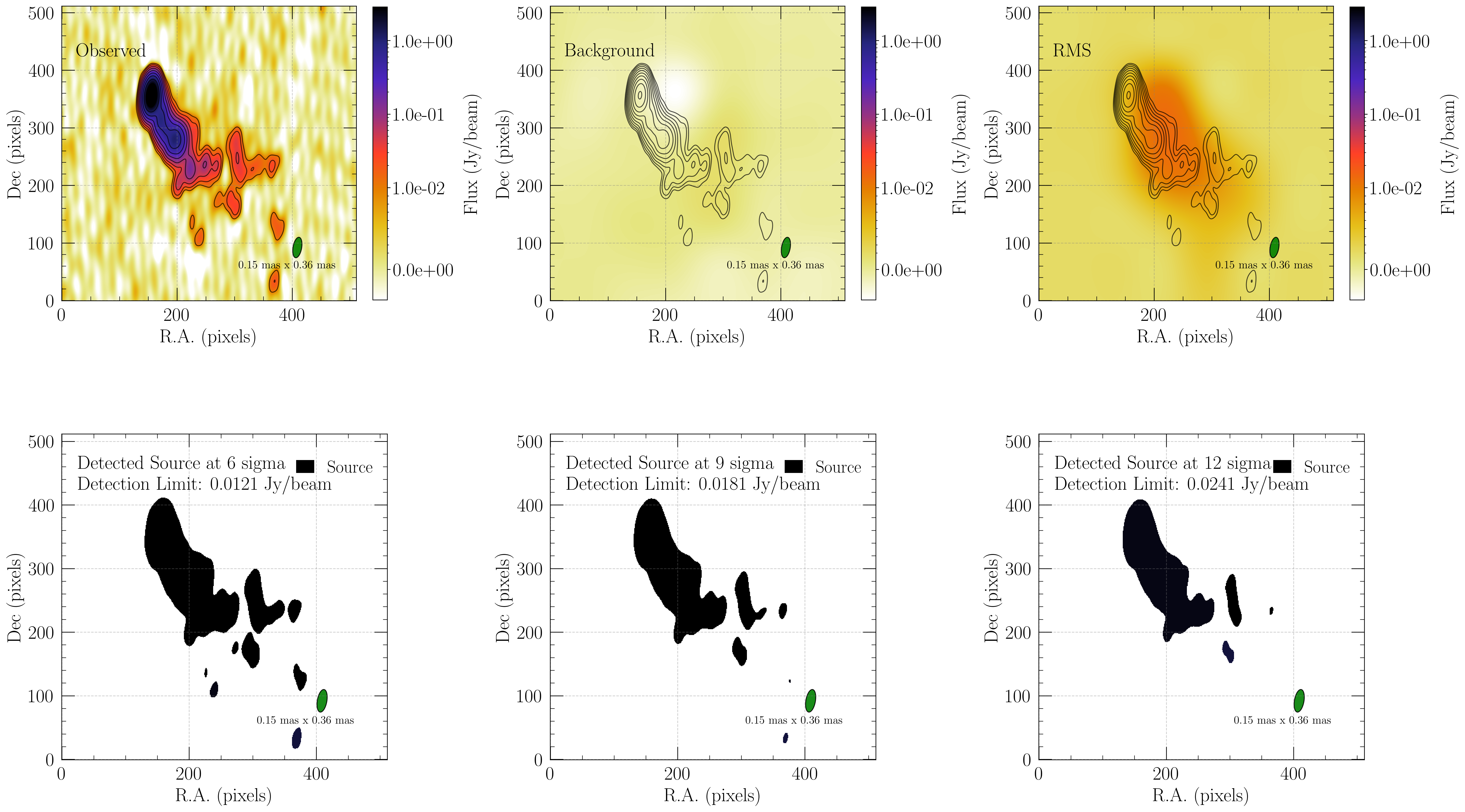}
\caption{Examples of different detection limits calculated for the VLBI intensity map of the blazar source 3C 279 under the BEAM-ME monitoring program. Top row: Observed map (Left) taken at 43 GHz on 2015 August 1, convolved with a beam of $0.15\times0.36$ mas$^{2}$ and a position angle of $-10^{\circ}$, background map (Middle), and RMS map (Right) returned by the detection limit function. The grey contours correspond to contour levels of 0.2, 0.4, 0.7, 1.4, 2.9, 5.8, 11.5, 23.0, 46.0, and $92.1\%$ of the peak total intensity. Bottom row: Masked pixels from the observed map with the detection limit calculated to be 6 (Left), 9 (Middle), and 12 (Right) times the median RMS. Pixels with values greater than the detection limit are masked with a value of 1, while pixels with values lower than the detection limit are set to be 0.}
\label{fig:detectionlimits}
\end{figure*}

\subsection{Detection Limit Estimation}

As mentioned in \autoref{sec:intro}, a key problem in component characterization in VLBI maps is the lack of clear edges between the source of interest and the background noise, especially with diffuse flux distributions. Without a well-defined detection limit, the component-fitting process may become susceptible to fitting noise peaks or artifacts, leading to spurious or misleading results. A detection limit ensures that only statistically significant features above a certain threshold are considered for component fitting. Also, a detection limit prevents overfitting, which could lead to artificially complex models that do not accurately reflect the underlying astrophysical emission.

To separate genuine emission from background noise in the intensity maps, the noise level is estimated using the \texttt{Background2D} class from the \texttt{photutils} package\footnote{\url{https://photutils.readthedocs.io/en/stable/api/photutils.background.Background2D.html}}. This tool computes a two-dimensional background model, with \texttt{MMMBackground}, by dividing the map into sub-regions and calculating the median intensity within each. The resulting output is an RMS map that characterizes the spatial variation of the background noise across the image.

To ensure a robust estimation of the background, we applied sigma clipping to exclude outlier pixels that may be affected by strong sources or imaging artifacts. We used the \texttt{SigmaClip} function from \texttt{astropy.stats}\footnote{\url{https://docs.astropy.org/en/stable/api/astropy.stats.SigmaClip.html}}, which iteratively rejects pixel values that deviate from the central value by more than a specified number of standard deviations. In our case, we adopted an asymmetric clipping threshold: pixels below $3\sigma$ and above $5\sigma$ from the median were excluded. This choice is motivated by the typical noise distribution in VLBI maps, which is approximately Gaussian but can contain bright outliers. The lower threshold removes faint artifacts and spurious noise, while the higher threshold prevents extremely bright features from biasing the background estimation. The sigma clipping is applied iteratively, recalculating the standard deviation of the remaining pixels at each step, until the process converges. This yields a stable and reliable RMS map that reflects the intrinsic noise structure of the data.

\texttt{Background2D} returns the 2D background model, 2D map of RMS values, and the median RMS value across the entire map. We define the detection limit as $\text{DL}=n\times\overline{\text{RMS}}$, where $\overline{\text{RMS}}$ is the median RMS, and $n$ is the threshold, an input parameter determined by the user, typically set to values between $6-12$, depending on the coverage of the visibility map of the source. Higher values of $n$ increase reliability by reducing false positives but could lose faint emission features. On the other hand, lower $n$ values will improve sensitivity but increase the risk of detecting noise and artifacts as signals, also increasing the computing time. An example of this can be seen on \autoref{fig:detectionlimits}, where for the same epoch map, different detection limits are established, leading to an alternative flux distribution to be fitted.

\subsection{Map Fitting}

Once the detection level is determined, the initial iteration takes place, one Gaussian is fitted at the position of the peak flux of the intensity map using \texttt{imfit}. The Gaussian function modeled on the region is parametrized by: i) Flux Density: Total flux associated with a component, ii) Position: In pixels, and iii) Orientation: Major axis, minor axis, and position angle (PA) of the elliptical Gaussian. The value for the first 2 parameters is the peak flux along with pixel coordinates in the map. The beam size will always be the initial guesses for the major ($a$) and minor ($b$) axes, and PA of the Gaussians fitted in each iteration. The reason for this arises from the fact that the beam in VLBI maps represents the instrumental resolution, which is usually an elliptical Gaussian. The beam defines the smallest spatial scale that can be resolved by the array. The major axis of the beam reflects the highest resolution along the primary direction of the synthesized beam, while the minor axis reflects the orthogonal shorter resolution. Consequently, any point source will appear smeared out, taking on a shape that approximates the dimensions and orientation of the beam. Because of this, and as a resolution proxy, the beam size provides a convenient initial estimate for the size and orientation of unresolved or partially resolved sources. 

The CASA task \texttt{imfit} employs an optimization routine, least-squares fitting, to refine these parameters and achieve the best fit to the observed intensity distribution, detailed description on \citet{McMullin2007}. Each iteration delivers the output of  \texttt{imfit}, i.e. the model and residual map (observed map minus model), along with the a log file with a detailed description of the properties of the fitted components, such as integrated flux and position, with its uncertainty.

After the initial iteration, the residual map is analyzed to check for significant leftover structures. If the peak flux in the residual map is higher than the detection limit, it means that significant flux associated with the source of interest are not modeled in the first iteration. Therefore, an additional Gaussian is added to the model, in the same position as the peak flux of the residual map, and then the fitting process is carried out again. This process is repeated until the peak flux of the residual map is below the detection limit. This iterative process ensures that multiple Gaussians are used if a single Gaussian is insufficient to fit complex structures. For cases where adding one more component to the fit causes the model to not converge, the algorithm stops and takes the previous iteration that converged as the final fit.

\subsection{Exclusion of Unreliable Components}

\begin{figure*}[t]
\centering
\includegraphics[width=2.1\columnwidth]{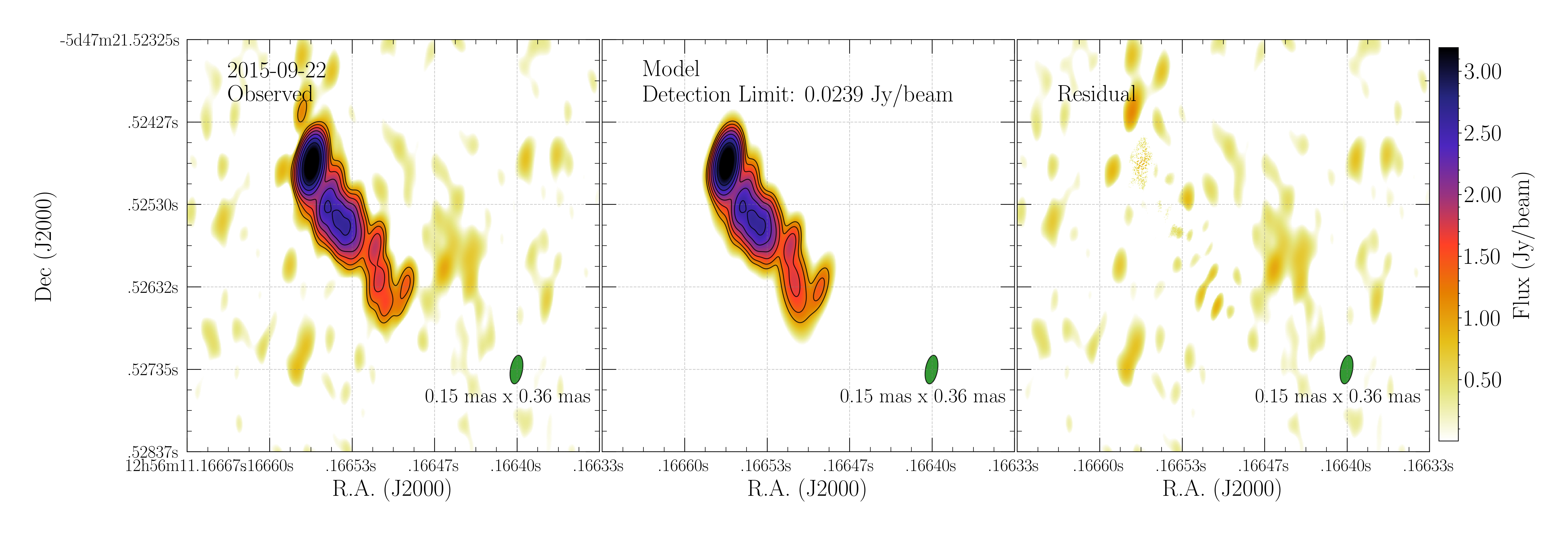}
\caption{Left panel: VLBI intensity map from the 3C 279 observation at 43 GHz in 2015 September 22, convolved with a beam of $0.15\times0.36$ mas$^{2}$ and a position angle of $-10^{\circ}$. Middle panel: Model of the observed flux distribution generated by the algorithm. Right panel: Residual image. The grey contours correspond to contour levels of 0.37, 0.75, 1.50, 3.00, 6.00, 12.00, 24.00, 47.93, and $95.90\%$ of the peak total intensity. The first contour level is at the detection limit.}
\label{fig:3C279}
\end{figure*}

To ensure the reliability of the fitted components, the routine incorporates constraints and checks on the Gaussian functions used, some fitted components may be unreliable due to low flux density, extreme eccentricity, or sizes smaller than the beam. These components are identified and removed from the fitting process after the first loop, to avoid introducing artifacts into the model. Criteria for component rejection:
\begin{itemize}
\item Detection Limit: Components with a peak intensity lower than this detection limit are discarded because they may be noise artifacts rather than real features
\item Elongated Components: Gaussians with high elongation are likely to reflect poorly constrained fits rather than real jet features. We apply a beam-relative criterion to rejecting any component whose major-to-minor axis ratio exceeds $f$ times the beam major-to-minor axis ratio, i.e. $\frac{a}{b}\geq f(\frac{a_{beam}}{b_{beam}})$. We adopt $f=3$, which for the average beam in our sample ($a/b\sim2.4$) corresponds to a threshold of $a/b\approx7$.
\item Size Constraint ($A_{\text{Gaussian}}<A_{\text{beam}}$): Components with area lower than the beam area are rejected because they fall below the resolution limit of the instrument and may not be physically meaningful.
\end{itemize}

After the remotion of unreliable components the filtered input model parameters are then fitted to the map. Once we re-run \texttt{imfit}, if the fitting process has converged, the result is exported as final. On the contrary, if the fitting was not successful, the function defaults back to the previous fitting results, indicating to the user that the remotion of doubtful components were unsuccessful and exporting the last fitting as final. In the case that no unreliable Gaussians were found, the function assumes the fitting results are already trustworthy and retains them, exporting the last iteration as final.

\subsection{Data Export}

For each fitted map, the output of the routine for every iteration (including the last one, labelled as final) are the usual outputs of CASA task \texttt{imfit}, which encompass the \texttt{txt} files: log, fit results, and summary, and \texttt{FITS/IMAP} files of the model and residual. In addition, a summary of the properties of the Gaussians fitted is delivered, for every fitted map, which contains the astrometric (position) and photometric (flux density) parameters, among others. The format of this component summary file is described in \autoref{tab:summary}. These results provide valuable information about the structure and properties of the radio-emitting regions in the VLBI map. Finally, an additional final log for all the maps fitted, in a machine-readable format, is provided with the information on the detection limit, number of components required, status of the last iteration (if the fit converges or not), status of the remotion of bad components (whether it was required or not, or if it was performed but failed to converge), RMS estimation of the map, final model, and final residual. Two additional parameters are also exported onto the summary; the rate between the residual RMS and the map RMS and the rate between the model RMS and the map RMS, called Residual Rate, and Model Rate, respectively. If the model accurately describes the flux distribution of the map, Residual Rate and Model Rate would be close to zero and one, respectively.

\begin{table}[t]
\caption{Summary component format.}
\label{tab:summary}
\begin{tabularx}{\columnwidth}{ccc}
\toprule
Column & Property & Units \\
\hline
1 & Integrated Flux & Jy \\
2 & Uncertainty in the Integrated Flux & Jy \\
3 & Peak Flux & Jy/beam \\
4 & Uncertainty in the Peak Flux & Jy/beam \\
5 & Right Ascension (J2000.0) & degrees \\
6 & Declination (J2000.0) & degrees \\
7 & Pixel position in x-axis & pixel \\
8 & Pixel position in y-axis & pixel \\
9 & Major-axis (FWHM) of the Gaussian & arcsec \\
10 & Uncertainty in the Major-axis & arcsec \\
11 & Minor-axis (FWHM) of the Gaussian & arcsec \\
12 & Uncertainty in the Minor-axis & arcsec \\
13 & Eccentricity of the Gaussian & no-units \\
14 & Position angle of the Gaussian & degrees \\
15 & Uncertainty in the Position angle & degrees \\
16 & Observed Frequency of the map & GHz \\
\hline
\end{tabularx}
\end{table}

\section{Testing}\label{sec:testing}

\begin{figure*}[t]
\centering
\includegraphics[width=2.1\columnwidth]{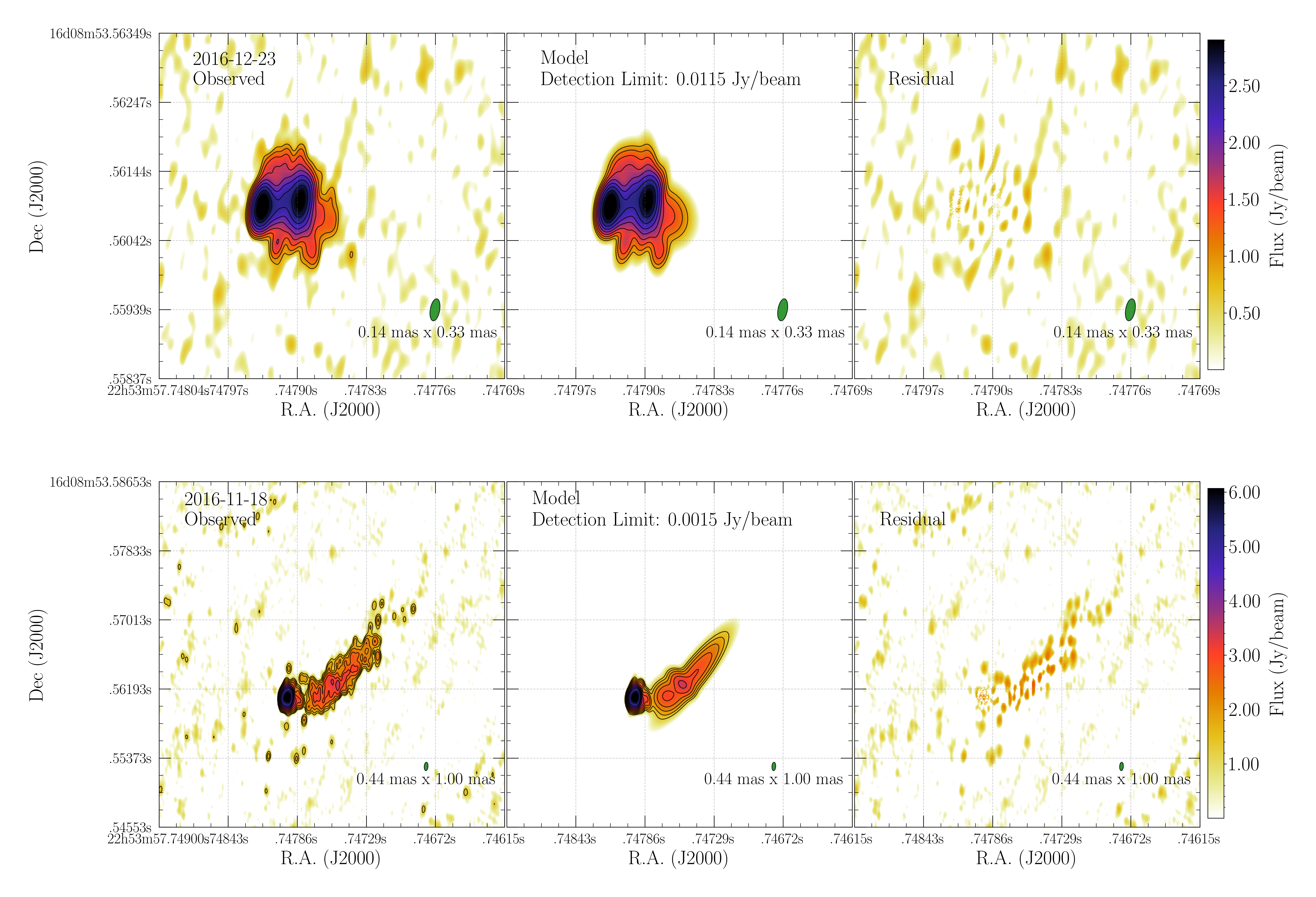}
\caption{Upper row: Left panel: VLBI intensity map from the 3C 454.3 observation at 43 GHZ in 2016 December 23, convolved with a beam of $0.14\times0.33$ mas$^{2}$ and a position angle of $-10^{\circ}$. Middle panel: Model of the observed flux distribution generated by the algorithm. Right panel: Residual image. The grey contours correspond to contour levels of 0.20, 0.40, 0.79, 1.60, 3.17, 6.40, 12.68, 25.36, 50.72, and $99.00\%$ of the peak total intensity. Bottom row: Similar as above Left panel: VLBI intensity map from the 3C 454.3 observation at 15 GHz in 2016 August 09, convolved with a beam of $0.44\times1.00$ mas$^{2}$ and PA$=-3.75^{\circ}$. Middle panel: Model of the observed flux distribution delivered by the algorithm. Right panel: Residual image. The grey contours correspond to contour levels of 0.02, 0.05, 0.10, 0.21, 0.41, 0.83, 1.66, 3.32, 6.63, 13.27, and $36.73\%$ of the peak total intensity. The first contour level at both observations is at the detection limit.}
\label{fig:3C454}
\end{figure*}

As a demonstration of the AAP-Imfit capabilities we present here the results from an end-to-end fitting of a representative VLBI dataset from blazar sources, part of the BEAM-ME and MOJAVE long-term monitoring programs. The BEAM-ME\footnote{\url{https://www.bu.edu/blazars/BEAM-ME.html}} (\citealp{Jorstad2017}) observations were carried out at 43 GHz (7 mm) and are also part of the Boston University VLBA Blazar monitoring program (VLBA-BU-BLAZAR). MOJAVE\footnote{\url{https://www.cv.nrao.edu/MOJAVE/project.html}}(\citealp{Lister2009}; \citeyear{Lister2013, Lister2018}) observations were taken at 15 GHz (2 cm). We test the reliability of this algorithm by comparing the automatically fitted maps with the ones used in \citet{PatinoAlvarez2019} and \citet{Palafox2025}, since in these works, they already fitted the VLBI maps for the sources 3C 279 \citep{PatinoAlvarez2019} and 3C 454.3 \citep{Palafox2025}. For both surveys we are using maps that cover a time period from 2008 to around 2020.

\subsection{3C 279 Fitting}

\begin{figure*}[t]
\centering
\includegraphics[width=2.1\columnwidth]{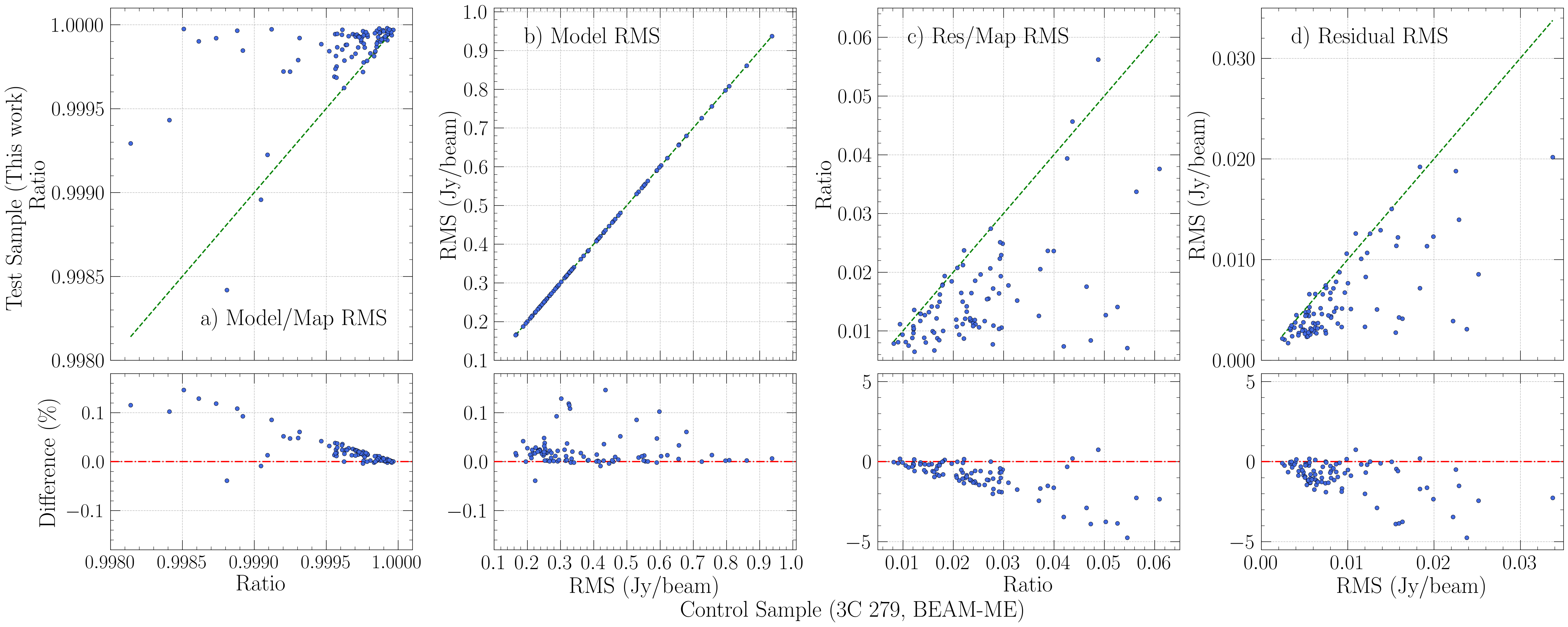}
\caption{Comparison between RMS values for the source 3C 279 using the observations from the BEAM-ME survey. (a) Upper: Rate between the model and map RMS, close values to one are expected if the model accurately retrieves the original flux distribution. Bottom: Percentage differences between both rates. (b) Upper: Model RMS, similar values between samples are expected if the routine retrieves similar models to the control sample. Bottom: Percentage differences between both model RMS values with respect to the observed map RMS. (c) Upper: Rate between the residual and map RMS, values close to zero are expected if the residual does not show source information and only background info. Bottom: Percentage differences between both rates. (d) Upper: Residual RMS, similar values between samples are expected if the routine retrieves a similar residual to the control sample. Bottom: Percentage differences between both residual RMS values with respect to the map RMS. Green dashed lines represent the linear one-to-one relation. The Red dashed line denotes zero difference.}
\label{fig:3C279RMS}
\end{figure*}

3C 279, cataloged as a flat spectrum radio quasar (FSRQ) at a redshift of $z=0.536$ \citep{Marziani1996}, is a highly variable radio source. \citet{PatinoAlvarez2019} linked $\gamma$-ray variability to a moving region downstream of the jet in the MOJAVE maps, constraining this emission zone to approximately 42 pc from the 15 GHz radio core. Furthermore, \citet{Okino2022} demonstrated that the jet in 3C 279 transitions from a parabolically collimated inner jet to an outer jet with canonical expansion, with the jet collimation break occurring around $10^{7}$ Schwarzschild radii.

As mentioned in \citet{PatinoAlvarez2019}, the component downstream of the jet in the MOJAVE maps was not modeled by Gaussian fitting, it was rather measured by manual integration of the flux region. Therefore to be consistent between methodologies and be able to compare homogeneously between the map models, we only use the 43 GHz BEAM-ME maps, which were modeled by Gaussian fitting in their entirety. We fit 92 maps that cover the period from 2008 to 2017, we use a detection limit of 8 times the median RMS. On average, the maps were fitted using 10 components. In \autoref{fig:3C279} are shown the observed, model, and residual map.

\subsection{3C 454.3 Fitting}

The FSRQ 3C 454.3, located at a redshift of $z=0.859$ \citep{JacksonAndBrowne1991}, ranks among the brightest sources in the $\gamma$-ray sky and is extensively monitored across various wavelengths, including VLBI (e.g. \citealp{Jorstad2013, Jorstad2017}). Its jet structure is complex, with evidence suggesting the presence of standing conical shocks. \citet{Jorstad2010} found that optical flares coincide with the passage of superluminal knots through the radio core, situated at the end of the collimation zone. \citet{Traianou2024} proposed a jet-bending model, where the relativistic plasma is nearly aligned with our line of sight, to explain the disappearance of jet features at 43 and 86 GHz. \citet{Palafox2025} identified multiple $\gamma$-ray emission zones: the radio core at 15 GHz and 43 GHz, and a moving region located downstream in the jet that propagates at an apparent velocity nearly ten times the speed of light.

We analyzed maps from the BEAM-ME and MOJAVE monitoring programs. For BEAM-ME observations, we fitted 131 maps spanning from 2008 to 2020, using a detection limit of 6 times the median RMS noise, requiring on average, 14 components per map. For the MOJAVE maps, we fitted 51 maps covering from 2008 to 2022, excluding those from 2019 due to a systematic flux calibration issue\footnote{Details in: \url{https://science.nrao.edu/enews/14.4/index.shtml/vlba_flux}}. A higher detection limit of 9 times the median RMS noise was used for this dataset, resulting in an average of 15 components per map. \autoref{fig:3C454} shows an example observed, model, and residual maps from both BEAM-ME and MOJAVE observations.

\section{Comparison with Control Sample}

\begin{figure*}[t]
\centering
\includegraphics[width=2.1\columnwidth]{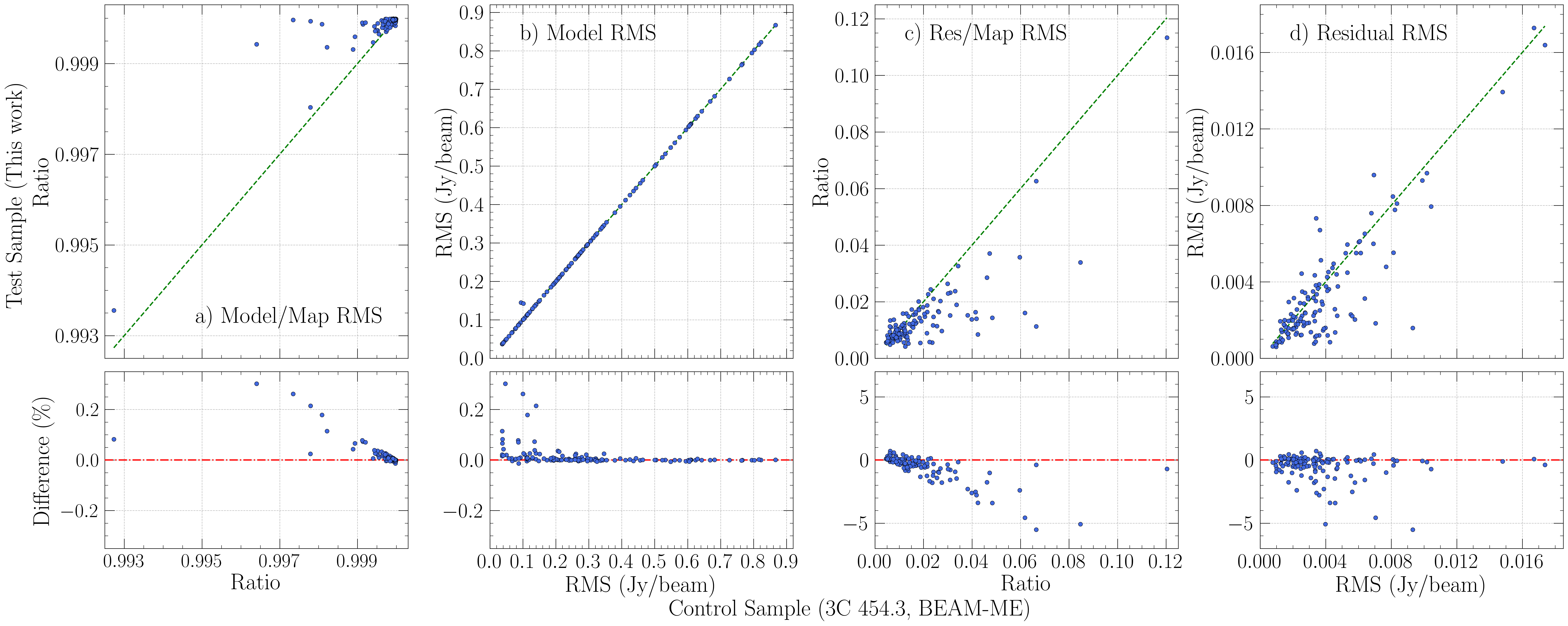}
\caption{Similar to \autoref{fig:3C279RMS}, but for the BEAM-ME observations of 3C 454.3.}
\label{fig:3C454RMSB}
\end{figure*}

\begin{figure*}[t]
\centering
\includegraphics[width=2.1\columnwidth]{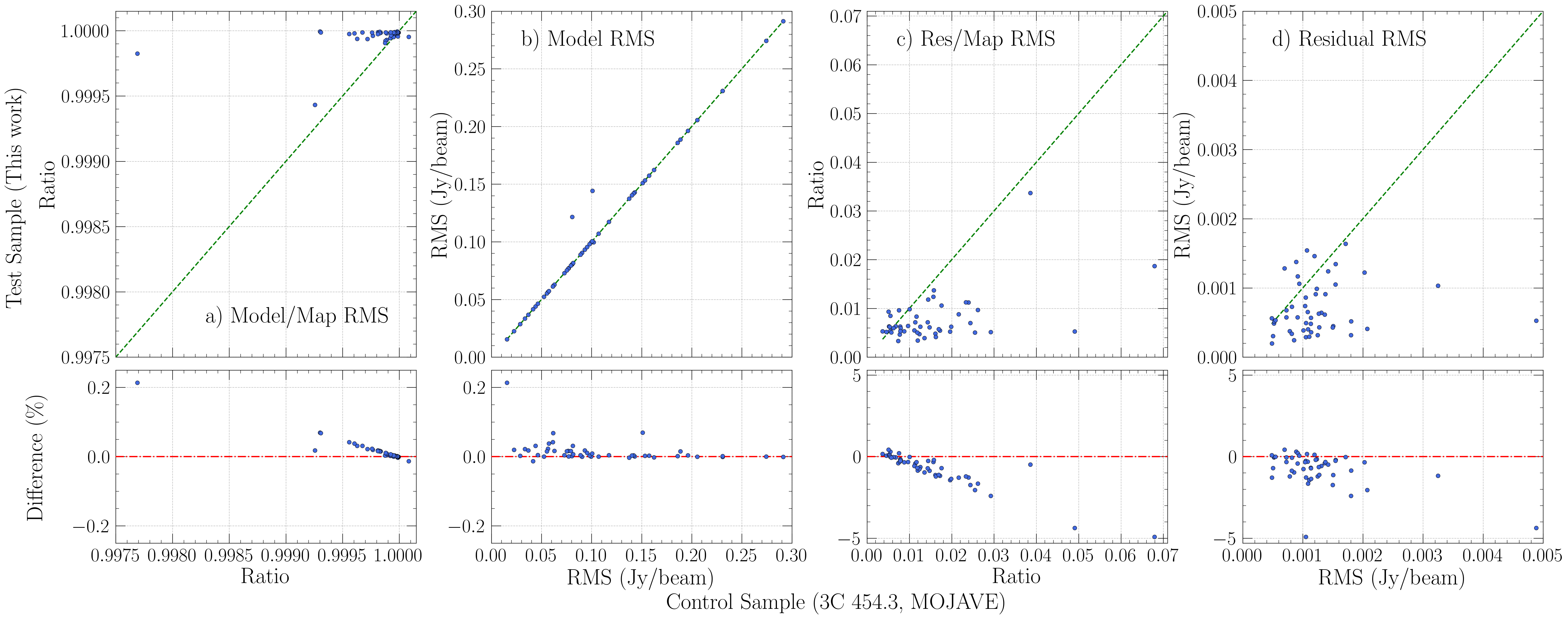}
\caption{Similar to \autoref{fig:3C279RMS}, but for the MOJAVE observations of 3C 454.3.}
\label{fig:3C454RMSM}
\end{figure*}

In order to test how well the automated fitting routine works, we use the datasets described in Section 3, specifically, we use the Root Mean Square as a metric of comparison. Through private communication, we obtained the model maps used in  \citet{PatinoAlvarez2019} and \citet{Palafox2025}, referred as the control sample. From these, we estimated the residual map and all the aforementioned RMS estimators.

The Model-to-Map ratio (Model/Map) directly compares the fitted model to the observed map, aiding in the detection of intensity over- or underestimation in specific regions, as well as sub- or overfitting of the observed map. A significant deviation from unity suggests the model inadequately retrieves the original flux distribution. Specifically, a ratio below 1 indicates a sub-fitted model, while values above 1 imply overfitting. The RMS of the fitted model quantifies the overall intensity variation in the reconstructed map. A significant difference between the RMS of the automatic model and that of the control sample may point to discrepancies in reconstructed intensity levels.

\begin{figure*}[t]
\centering
\includegraphics[width=2.1\columnwidth]{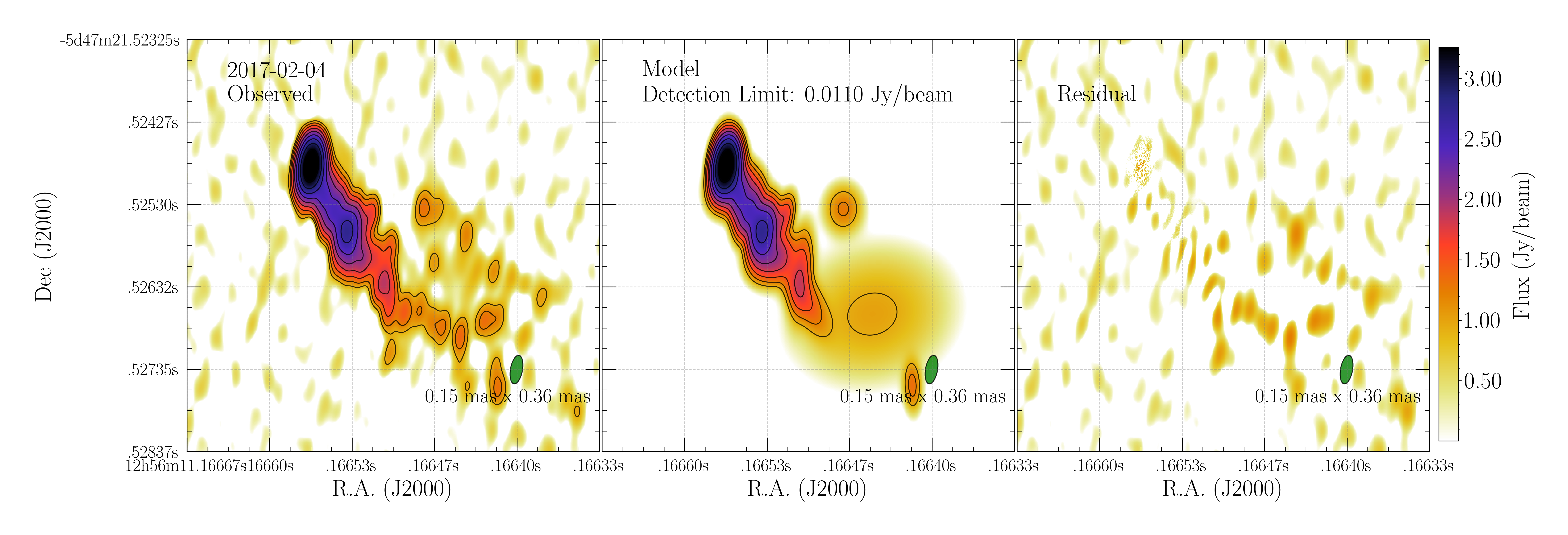}
\caption{Example of model delivered by the routine where a manual improvement could be performed. Left panel: VLBI intensity map from the 3C 279 observation at 43 GHz in 2017 February 4, convolved with a beam of $0.15\times0.36$ mas$^{2}$ and a position angle of $-10^{\circ}$. Middle panel: Model of the observed flux distribution generated by the algorithm. Right panel: Residual image. The grey contours correspond to contour levels of 0.37, 0.75, 1.50, 3.00, 6.00, 12.00, 24.00, 47.93, and $95.90\%$ of the peak total intensity. The first contour level is at the detection limit.}
\label{fig:3C279improve}
\end{figure*}

The Residual-to-Map ratio (Res/Map) normalizes the residuals by the intensity of the observed map, making it useful for identifying regions with the largest relative model deviations. High values of this ratio indicate areas where the model fails to accurately reconstruct the observed structure. We anticipate values close to zero in most comparisons, signifying small residuals relative to the observed intensities. Finally, the residual map, calculated as the difference between observed - fitted maps, highlights areas of disagreement. The RMS of the residual map quantifies the overall deviation between the model and the observed map. A low residual RMS indicates a good fit, whereas a high residual RMS suggests systematic differences or missing structures in the model.

\autoref{fig:3C279RMS}, \autoref{fig:3C454RMSB}, and \autoref{fig:3C454RMSM} present comparisons of the RMS values for the rates, models, and residuals among our fitted maps, the test sample, and the control samples from either \citet{PatinoAlvarez2019} or \citet{Palafox2025}. In all the aforementioned figures, the lower panels display the differences (in percentages) in the RMS values between the test and control samples. A positive y-axis value means the RMS from the test sample is larger than that of the control sample, indicating that the control model provides the better fit. Conversely, a negative value means the RMS from the test sample is smaller, demonstrating that the automated routine produces the more accurate fit. For the 3C 279 Fits (\autoref{fig:3C279RMS}):

\begin{itemize}
\item The Model-to-Map RMS ratio is slightly higher for the test sample, with differences around $0.15\%$ compared to the observed map RMS.
\item Model RMS values exhibit a near one-to-one relationship with the control sample, showing similar deviations of approximately $0.15\%$.
\item Conversely, the Residual-to-Map ratio and the residual RMS distribution show slightly larger values for the control sample, with differences under $5\%$ relative to the observed map RMS. A number of points exceed 0.02; which indicate instances where the control sample models under-fit the data, i.e. the automated routine provides a more accurate fit than the control sample, with the residuals being up to $5\%$ lower.
\item AAP-Imfit models retain a flux distribution similar to the models of the control sample, with differences in the order of $0.15\%$. This slight overestimation likely arises because the automatic routine tends to fit several small Gaussians to extended emission along the jet, capturing specific flux peaks, whereas the control sample uses fewer, larger Gaussians that smooth over these peaks.
\item This approach results in smoother residuals in our fits, as evidenced by the approximately $5\%$ difference in the residual RMS values between the test and control sample. The majority of 3C 279 maps in our sample do not exhibit complex extended emission, leading to overall well-fitted models.
\item Visual inspection suggests that manual improvement could be beneficial for around 15 models (epochs) where diffuse flux distributions are present. In these cases, the use of a larger, smoother Gaussian was fitted to represent the overall structure, albeit at the cost of losing some specific flux details. An example of this scenario is showed on \autoref{fig:3C279improve}.
\end{itemize}
Similar RMS values distributions are observed for 3C 454.3 in both the BEAM-ME (\autoref{fig:3C454RMSB}) and MOJAVE (\autoref{fig:3C454RMSM}) maps. On average, the difference in the model RMS values between samples is $0.2\%$, while for the residual RMS values the difference is approximately $3\%$. However, this source exhibits some outlier data points.

\begin{figure}[h]
\centering
\includegraphics[width=\columnwidth]{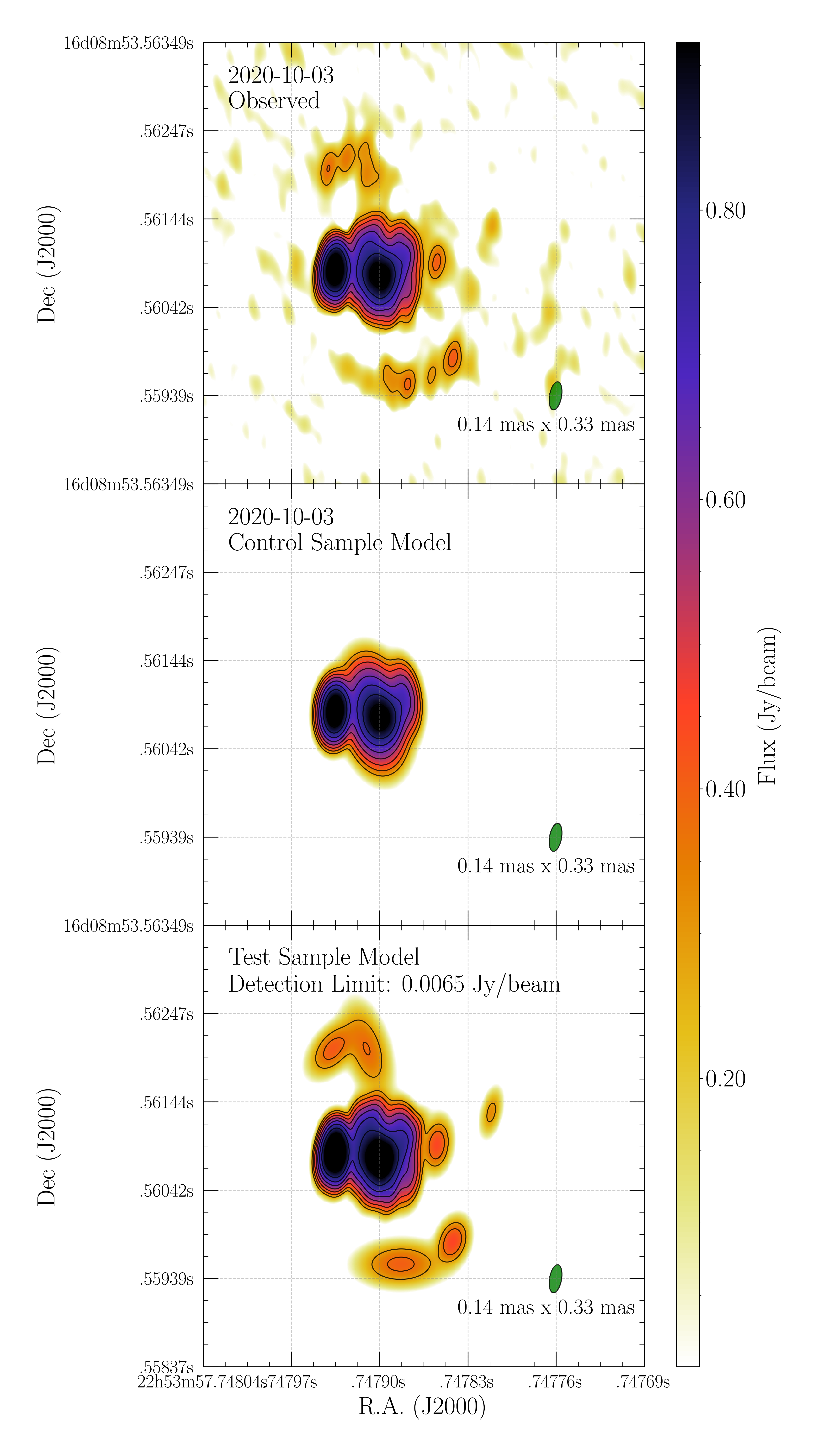}
\caption{Upper panel: VLBI intensity map from the 3C 454.3 observation at 43 GHZ in 2020 October 3, convolved with a beam of $0.14\times0.33$ mas$^{2}$ and a position angle of $-10^{\circ}$. Middle Panel: Corresponding retrieved model from \citet{Palafox2025} Bottom panel: Corresponding model of the observed flux distribution generated by the routine. The first contour level is at the detection limit.}
\label{fig:3C454bad}
\end{figure}

BEAM-ME Maps: Two epochs show significant deviations from the one-to-one relationship in model RMS values (see \autoref{fig:3C454RMSB}b), indicating substantial differences between our fitted model and the control sample model for those epochs. This discrepancy arises because our model fits artifacts of the interference pattern with high apparent flux, exceeding the detection limit (see \autoref{fig:3C454bad}). Consequently, the Model-to-Map RMS ratio for these two epochs is greater than unity (excluded from \autoref{fig:3C454RMSB}a for better visualization), representing a clear case of overfitting with a ratio difference exceeding $40\%$. Furthermore, the flux density of our models is higher than that of the corresponding observed maps, indicating a failure of the automatic routine to accurately retrieve component photometry for these instances. Nevertheless, the availability of each fitting iteration allows for potential model improvement at earlier stages.

\begin{figure}[t]
\centering
\includegraphics[width=\columnwidth]{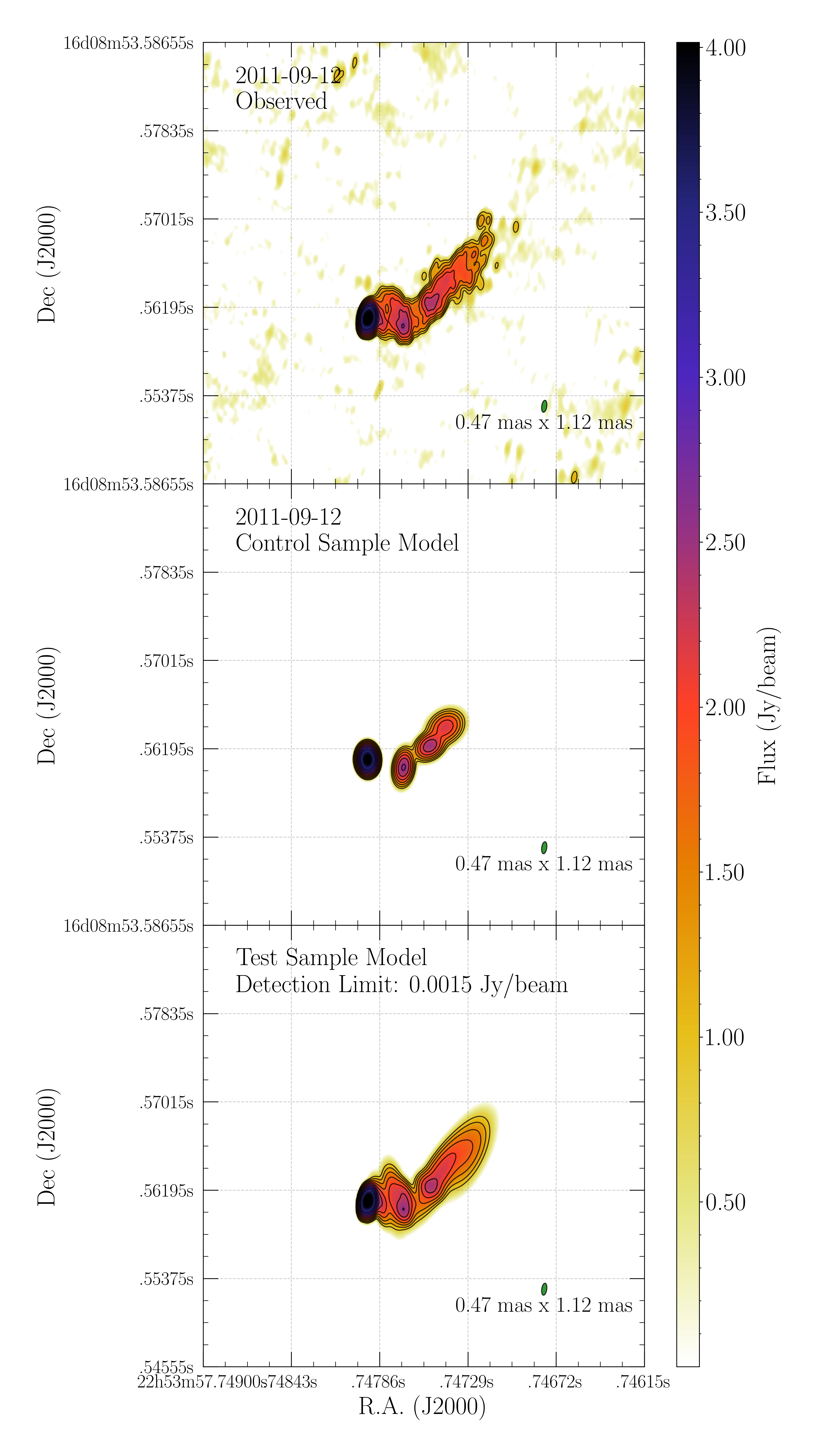}
\caption{Upper panel: VLBI intensity map from the 3C 454.3 observation at 15 GHz in 2011 September 12,, convolved with a beam of $0.53\times1.03$ mas$^{2}$ and a position angle of $0.53^{\circ}$. Middle Panel: Corresponding retrieved model from  \citet{Palafox2025}. Bottom panel: Corresponding model of the observed flux distribution generated by the routine. The first contour level is at the detection limit.}
\label{fig:3C454good}
\end{figure}

3C 454.3 displays a more complex flux distributions, with several maps showing diffuse structures. Some of these diffuse regions are fitted with a single extended Gaussian with low integrated flux in the control sample models. Out of 131 fitted maps, approximately 45 epochs could be improved by either eliminating components fitting artifacts or using multiple Gaussians to better represent specific regions.

MOJAVE Maps: Due to the large size of the maps compared to the source, we wanted to avoid a disproportional amount of background pixels (with respect to source pixels), because this would skew the RMS value towards the background instead of the source flux distribution. To mitigate this effect in the MOJAVE survey, we cropped the observed, model, and residual maps for both our sample and the control sample. We estimate the RMS from only a fraction of the map, the cropped section, the largest common area containing non-zero pixels in either model. It is worth to mention that for 3C 454.3 the flux distribution in MOJAVE maps often appears more detailed or ``granular''. This granularity complicates Gaussian fitting, as previously mentioned, fewer, larger, smoother Gaussians were fitted to represent the overall structure rather than fitting many smaller Gaussians. Manual improvement could be made in these cases,, an example of this is shown in the lower residual panel of \autoref{fig:3C454}.

The residual features arise due to the ‘granular’ flux distribution in the MOJAVE maps, which leads to under-fitting with the automatic Gaussian components. This situation is typical and could be improved by manually adding smaller Gaussians in regions of high flux granularity.

Similar to the BEAM-ME results, the model RMS values for two epochs deviate from the one-to-one relation (see \autoref{fig:3C454RMSM}b), again occurring when the models differ substantially. In these cases, the Model-to-Map RMS ratio is smaller for the control sample ($\leq0.7$), while our test sample ratio is around 1, this implies a case where the model delivered by the automatic routine represents better the flux distribution of the observed map (excluded from \autoref{fig:3C454RMSM}a for better visualization). \autoref{fig:3C454good} show the models comparison for one of the epochs in question. For this epoch the improvement even translates to the residual RMS and its corresponding ratio, which show a higher value in the control sample ($>0.2$, also excluded from \autoref{fig:3C454RMSM}d for better visualization).

In general, AAP-Imfit accurately retrieves the observed flux distribution for most maps, significantly reducing manual fitting time. However, the final fits should be considered preliminary until visual inspection, as manual refinement may be necessary. Specifically, modeling components smaller than the beam size, or diffuse extended regions poses a challenge for the routine. Since each iteration is provided, users can decide whether to refine the model from an earlier stage.

\section{Application to Astronomical Images}\label{sec:applications}  

One of the most important applications of this approach lies in studying flux variability within AGN jet components. Blazars and other relativistic jet sources often exhibit rapid flux changes due to intrinsic variability, relativistic boosting, and interactions with the surrounding medium (e.g., \citealp{MarscherAndGear1985, Aller1985, Hovatta2008}). By fitting each component separately and tracking their photometry across multiple epochs, the automatic routine provides a precise method for measuring flux evolution at milliarcsecond resolution. This is particularly useful for distinguishing between variations in a source total flux and structural changes within the jet, which can alter the observed flux distribution \citep{Jorstad2005, Rani2015, Okino2022}. The automation of this process ensures consistency across epochs, avoiding subjective biases inherent in manual component selection.

Furthermore, retrieving the flux of individual components enables the investigation of energy dissipation and particle acceleration processes in AGN jets. Synchrotron emission from relativistic electrons is expected to evolve due to cooling mechanisms affecting lower-frequency emission, shock interactions, and magnetic field variations \citep{MarscherAndGear1985, Urry1997, BaiAndLee2003, Fromm2013}. Analyzing the photometry of distinct jet components over time allows for testing theoretical models of energy loss and jet dynamics. The algorithm facilitates a quantitative assessment of how flux variations correlate with component motion, providing insights into shock formation, turbulence, magnetic reconnection events in jets \citep{Marscher2008, Jorstad2017}, and even with $\gamma$-ray flux variability. This last point is a key motivation for developing this algorithm: correlating the flux variability of individual VLBI components with simultaneous $\gamma$-ray flux. This method can determine if a percentage of VLBI variability in a specific component is related to $\gamma$-ray flux variability, thus pinpointing the $\gamma$-ray emission zone not only to different jet regions but also to specific observed components and their evolution \citep{PatinoAlvarez2019, Palafox2025}. The automatic nature of the code ensures efficient execution of these analyses, even for large VLBI datasets.

Another significant advantage of component-based photometric analysis is its role in studying Doppler boosting effects. Since relativistic beaming can significantly alter the observed flux of jet components, tracking the photometric evolution of individual features allows for better constraints on their velocity and orientation \citep{Sher1968, Lister2009}. This can refine estimates of jet speeds and viewing angles, improving our understanding of the connection between observed variability and jet kinematics. Unlike total flux monitoring, which integrates emission from multiple unresolved structures, a component-wise approach provides a clearer picture of individual region contributions to the overall emission. The automation of this process ensures consistent tracking of flux variations across epochs, enhancing the accuracy of variability studies.

Beyond AGN studies, the ability to retrieve individual component fluxes has applications in other areas of high-resolution astrophysics. For instance, in VLBI observations of gravitationally lensed systems, monitoring the photometric evolution of multiple lensed images can provide constraints on microlensing effects and substructure within the lensing galaxy \citep{Biggs2003, Blackburne2011}. It is important to note, however, that the observed maps are convolved with the beam, potentially leading to more complex flux distributions due to the nature of gravitational lensing. Similarly, in studies of radio supernovae and transient events, resolving and tracking the flux of expanding shock fronts can yield valuable information about explosion dynamics and circumstellar environments \citep{Bietenholz2003, Marcaide2009}. In this context, \citet{MartiVidal2024} introduced a Markov-chain Monte Carlo approach to VLBI imaging and source centering for SN1993J, which yields highly reliable maps. Since our routine, operates on already reduced intensity maps, both methods are complementary: the Markov-chain technique is ideal for generating high-fidelity maps in critical cases, whereas our algorithm is optimized for the subsequent stage of large-scale component characterization. When maximum accuracy is required, re-imaging with advanced techniques should precede automated fitting.

Overall, the main strength of this routine lies in its ability to provide high-precision photometric measurements at VLBI scales in an automated and efficient manner, enabling detailed studies of flux variability, jet dynamics, and Doppler boosting effects. While astrometry remains an essential aspect of VLBI image analysis, the capacity to systematically extract and monitor the flux of individual jet components represents a significant advantage for studying the physical processes that govern compact radio sources. Applying this technique to large datasets makes it possible to uncover statistical trends in jet evolution and variability, contributing to a deeper understanding of AGN physics and high-energy astrophysical phenomena.

\section{Conclusions}

The automated routine developed in this work introduces a novel approach to analyzing VLBI intensity maps by fitting individual components and retrieving both astrometry and photometry. While astrometric measurements are obtainable through other established methods, the key advantage of this algorithm lies in its systematic extraction of individual component flux densities, enabling detailed studies of their variability and evolution. The public version of AAP-Imfit is available as a GitHub/Zenodo repository\footnote{ \url{https://github.com/Alfred97AstroAGN/AAP-Imfit-aCASA-tool.git}} \citep{AAPimfit2025}, with a small number of VLBI intensity maps retrieved from the test sample, see \autoref{sec:testing}. Despite the time-saving benefits, the automated fitting process has limitations. The complexity of flux distributions in VLBI maps, particularly in sources like 3C 454.3, can pose challenges. Faint emission regions or structures smaller than the beam size can be difficult for the algorithm to model accurately. Although the code produces a fully fitted map above the detection limit, visual inspection remains necessary to ensure a physically meaningful fit, and manual refinement may occasionally be required. The provision of each fitting iteration offers user flexibility to resume from the final fit or revert to a previous one based on the quality of the results. Therefore, we recommend using this routine as a tool to minimize fitting time, rather than as a definitive fit. This algorithm achieves two key outcomes:

\begin{itemize}
\item Reproducing the flux distribution of VLBI intensity maps: The algorithm accurately models the observed flux distribution, as evidenced by the small RMS of the residual maps, the similarity in RMS values between our fits and control samples, and the Model-to-Map and Residual-to-Map ratios approaching one and zero, respectively.
\item Enabling large-scale VLBI component characterization: By automating the fitting process, this algorithm facilitates the analysis of significantly larger VLBI map samples. Previously manual, component characterization is now streamlined and accelerated, substantially reducing the time required for large datasets and making their analysis feasible.
\end{itemize}

While this automation significantly enhances efficiency and reproducibility, it presents computational demands, echoing limitations found by \citet{Caproni2011} regarding processing time with multiple Gaussians. The component fitting software, CASA task \texttt{imfit}, is not optimally designed for handling numerous simultaneous components, potentially increasing processing times, especially for large source samples across many epochs. Maximizing computational performance necessitates a system with sufficient RAM, as processing multiple components concurrently can be memory-intensive. While typical cases can run comfortably with 8-16 GB, large-scale analyses, especially those fitting dozens of components over many epochs, may require up to 32 GB or more to maintain responsiveness and avoid memory bottlenecks. Optimizing the workflow to minimize unnecessary computational overhead is crucial for extensive datasets. Future improvements, such as using alternatives to CASA imfit, could further reduce computational overhead and enhance scalability. Additionally, future code enhancements could include automatically grouping multiple Gaussians fitting a single region and tracing components across epochs for more effective evolution studies. These improvements would further empower the algorithm for systematic investigations of AGN jet structures and other astrophysical phenomena.

\section*{Acknowledgments}
We thank the anonymous referee for the constructive comments that helped to improve the manuscript. A.A.-P. gratefully acknowledges the support received from the SECIHTI (Secretaria de Ciencias, Humanidades, Tecnología e Inovación) program for their Ph.D. studies. E.P. acknowledges the financial support provided by the National System of Researchers (SNII - SECIHTI) through a research assistant scholarship (SNII III). S.A.D. acknowledges the M2FINDERS project from the European Research Council (ERC) under the European Union's Horizon 2020 research and innovation programme (grant No. 101018682). This research was made possible thanks to the assistance provided by the Max Planck Institute for Radio Astronomy (MPIfR) - Mexico Max Planck Partner Group led by V.M.P.-A. This research made use of Photutils, an Astropy package for detection and photometry of astronomical sources. This research has made use of data from the MOJAVE database that is maintained by the MOJAVE team. This study makes use of VLBA data from the VLBA-BU Blazar Monitoring Program (BEAM-ME and VLBA-BU-BLAZAR, funded by NASA through the Fermi Guest Investigator Program. The VLBA is an instrument of the National Radio Astronomy Observatory. The National Radio Astronomy Observatory is a facility of the National Science Foundation operated by Associated Universities, Inc.

\software{AAP-imfit \citep{AAPimfit2025}, Astropy \citep{Astropy2013, Astropy2018, Astropy2022}, CASA \citep{McMullin2007, CASA2022}, Photutils \citep{Phoutils2025}
          SciPy \citep{Virtanen2020}.
          }
          
\appendix
\section*{Software Requirements}\label{sec:req}
The automated component-fitting routine was developed in Python and requires the following packages: numpy 1.24.4, pandas 2.0.3, matplotlib 3.7.5, astropy 5.2.2, photutils 1.0.2, casatasks 6.6.5.31, and casatools 6.6.5.31. The code can be cloned from the GitHub repository using git clone \url{https://github.com/Alfred97AstroAGN/vlbi-auto-photometry-tool.git}, and dependencies can be installed either via \texttt{pip install -r requirements.txt} or manually using pip. The main fitting functions are implemented in \texttt{main\_functions.py} and can be called directly within a script or notebook. Two example notebooks are included: \texttt{main\_usage.ipynb}, which provides a detailed usage guide, and \texttt{demo.ipynb}, which demonstrates a basic application on a small VLBI map. Each map (e.g., .fits or .IMAP) should be stored in its own folder, and users must provide both a list of full paths to the files and a separate list of the parent directories. We strongly recommend fitting maps from a single source at a time to ensure a consistent detection limit, as the optimal threshold may vary between sources. The routine was tested in environments with at least 16–32 GB of RAM; for large-scale datasets with many components, 64 GB or more is recommended to ensure smooth performance.

\bibliography{AAP-Imfit_Manuscript}{}
\bibliographystyle{aasjournal}

\end{document}